\documentstyle[aps,prb]{revtex}

\begin{document}
\draft

\title{Electronic and phononic states of the Holstein-Hubbard 
dimer of variable length}

\author{M. Acquarone} 

\address{G.N.S.M.-C.N.R., Unit\`a I.N.F.M. di Parma, Dipartimento di
Fisica,\\ Universit\'a di Parma, 43100 Parma, Italy}

\author{J. R. Iglesias and M. A. Gusm\~ao}

\address{Instituto de F{\'\i}sica, Universidade Federal do Rio
Grande do Sul, 91501-970 Porto Alegre, RS, Brazil}

\author{C. Noce and A. Romano}

\address{Unit{\'a} I.N.F.M. di Salerno, Dipartimento di
Fisica Teorica e S.M.S.A.,\\ Universit{\'a} di Salerno, 84081
Baronissi, Salerno, Italy}

\maketitle

\begin{abstract}
 We consider a model Hamiltonian for a dimer of length $a$ including
all the electronic one- and two-body terms consistent with a single
orbital per site, a free Einstein phonon term for a frequency
$\Omega$, and an electron-phonon coupling $g_0$ of the Holstein type.
The bare electronic interaction parameters were evaluated in terms of
Wannier functions built from Gaussian atomic orbitals. An effective
polaronic Hamiltonian was obtained by an unrestricted
displaced-oscillator transformation, followed by evaluation of the
phononic terms over a squeezed-phonon variational wave function.  For
the cases of quarter-filled and half-filled orbital, and over a range
of dimer length values, the ground state for given $g_0$ and $\Omega$
was identified by simultaneously and independently optimizing the
orbital shape, the phonon displacement and the squeezing effect
strength.  As $a$ varies, we generally find discontinuous changes of
both electronic and phononic states, accompanied by an appreciable
renormalization of the effective electronic interactions across the
transitions, due to the equilibrium shape of the wave functions
strongly depending on the phononic regime and on the type of ground
state.
\end{abstract}



\section{Introduction}
\label{sec:intro}

Much attention has been devoted to materials where structural effects
deeply influence the electronic phenomena, as, for instance,
V$_{2-x}$Cr$_x$O$_3$,\cite{v2o3} polyacetilene,\cite{campbell,mukhop}
displacive-type ferroelectrics,\cite{ishi} the manganites exhibiting
colossal magnetoresistance,\cite{GMR} and the high temperature
superconducting perovskites.\cite{egami,bianconi,good} From the
theoretical standpoint, such effects have been studied by adding
various kinds of electron-phonon couplings to different electronic
Hamiltonians, mainly of the extended Hubbard or $t$--$J$ type.
However, comparatively little attention has been paid to the effect of
the structure (in its simplest form, a variation of the lattice
spacing $a$) onto the electronic interactions themselves.  Such
effects are of great relevance in experiments: consider, for example,
the measurements under either external or chemical pressure, and the
wealth of interesting phenomena that they have revealed. The
electronic interaction parameters can be written as integrals of
various combinations of Wannier functions centered on the lattice
sites (as will be detailed in Section \ref{sec:model} below).
Therefore, a change of $a$ is expected to have a non-negligible effect
on them.  In the realm of electronic transitions, to our knowledge,
after a pioneering attempt by Kawabata\cite{Kawabata}, only Spa\l ek
and W\'ojcik\cite{js92} focused onto such effects. They argued that,
in a system of fermions undergoing a metal-insulator transition (MIT)
due to the on-site repulsion effect, the wave functions may
significantly change their shape across the transition, concomitantly
to a change of the equilibrium lattice spacing. Therefore, different
values for the hopping amplitude $t_{ij}$ and the on-site repulsion
$U$ may correspond to the itinerant and the localized regimes.  In the
present work we explore the consequences of such an approach in
systems with a Holstein-type electron-phonon interaction. Our model
Hamiltonian includes all the electronic interaction parameters for
one- and two-body terms, whose dependence on $a$ results from their
evaluation in terms of Wannier functions built from Gaussian orbitals.
We also consider the characteristic structural length $a$ as free to
vary between reasonable boundaries, in order to model the effects of
the experimental techniques that allow for a more or less continuous
tuning of the bond length.  We find that both the electronic and
phononic states significantly depend on $a$.  For instance, on one
side, the phonon state, characterized by the extent of the {\it
displacement\/} and the {\it squeezing\/}, reflects the electronic
transitions at a deeper level than just a frequency renormalization.
On the other side, as the optimal shape of the Wannier wave functions
depends not only on $a$, but also on the phonon state, the
renormalization of the electronic parameters by the phonons is
actually more complex than indicated by the standard results of the
polaronic model.\cite{masc94}

 To put into evidence the basic features of our approach, avoiding
unnecessary complications, our model system in the present paper is a
simple dimer with a single orbital per site. It can be considered as
the basic unit to build clusters and lattices, and its simplicity
allows for a full control on the analytic aspects of the calculations.
Also, some unavoidable approximations which have to be introduced are
linked to the physics, and not to the complications of treating larger
systems.  Some new aspects distinguish the present work from the rich
existing literature on the Holstein dimer with various types of
inter-electronic interactions, (see, for instance, Refs.\
\onlinecite{campbell,Ranninger,wellein}, and the literature cited
therein). First, the effective electronic Hamiltonian originates from
a first-principles Hamiltonian including the complete set of one- and
two-body electronic terms consistent with a single orbital per site.
To the best of our knowledge, previous studies considered only a more
restricted subset of interactions. Second, we present our results
mainly by setting the values of the electron-phonon coupling $g_0$ and
phonon energy $\hbar\Omega$ and taking the dimer length $a$ as the
continuously varying controlling parameter.  The dependence of the
state of the system on its length usually receives little attention
even though it is much more easily accessible experimentally than the
dependence on the strength of the coupling, which requires delicate
chemical manipulations. However, some discussion of the case of fixed
length and varying coupling will also be given. We will discuss the
cases of quarter- and half-filling of the electron orbitals.

 The paper is organized as follows: Section \ref{sec:model} presents
the model Hamiltonian and the expressions of the electronic
interactions explicitly depending on the dimer length. In Section
\ref{sec:Heff} the effective electronic Hamiltonian is obtained, while
in Section \ref{sec:cf} several correlation functions and other
quantities of experimental interest are explicitly evaluated. The
results of the numerical analysis of our model are presented and
commented in Section \ref{sec:results}. Finally, some general
conclusions are presented in Section \ref{sec:concl}.

\section{The model}
\label{sec:model}

For a dimer consisting of identical ions centered at ${\bf R}_1$ and
${\bf R}_2$ we consider the Hamiltonian $H\equiv H^{}_{\text{el}}+
H^{0}_{\text{ph}}+H^{}_{\text{ep}}$ where $H_{\text{el}}$ includes all 
the electronic one-
and two-body terms for a single orbital, $H_{\text{ph}}^0$ is the free-phonon
term for a single Einstein frequency $\Omega$, and $H_{\text{ep}}$ introduces
an electron-phonon coupling $g_0$ of the Holstein type.

 In standard notation, we have
\begin{eqnarray}
\label{eq:hel}
H^{}_{\text{el}} & = &  \epsilon_0 \sum_\sigma (n^{}_{1\sigma} + 
n^{}_{2\sigma})
 - \sum_{\sigma} \left[ t  - X (n^{}_{1,- \sigma} + n^{}_{2,- \sigma})
 \right] (c^\dagger_{1\sigma} c^{}_{2\sigma} + H.c.)
 + U\sum_{i=1,2} n^{}_{i\uparrow} n^{}_{i\downarrow} \nonumber \\
&& \mbox{~} + V n^{}_1  n^{}_2  - 2J^{}_z S^z_1 S^z_2            
- J^{}_{xy}  (S^+_1 S^-_2 + H.c.)
+ P (c^\dagger_{1\uparrow}c^\dagger_{1\downarrow}
c^{}_{2\downarrow} c^{}_{2\uparrow} + H.c.) \;.
\end{eqnarray}

The bare electronic parameters $\epsilon_0$, $t$, $X$, $U$, $V$, and
$P$ ($= J_z = J_{xy}$) were evaluated according to Ref.\
\onlinecite{hubbard1}, using normalized Gaussian ``orbitals'' $\phi_i
({\bf r}) = \left(2 \Gamma^2/\pi \right)^{3/4} \exp\left[-\Gamma^2
({\bf r} - {\bf R}_i)^2 \right]$ ($i=1,2$). Their overlap integral is
$S \equiv \langle \phi_1 | \phi_2 \rangle = \exp(- \Gamma^2 a^2 /2)$,
where $a=|{\bf R}_1-{\bf R}_2|$ is the length of the dimer.  Then the
two orthonormalized Wannier functions are
\begin{eqnarray}   \label{eq:Wannier}
\Psi_1 ({\bf r}) &=& A_+(S) \phi_1 ({\bf r}) + A_-(S) \phi_2 ({\bf r})
 \;, \nonumber \\
\Psi_2 ({\bf r}) &=& A_-(S) \phi_1 ({\bf r}) + A_+(S) \phi_2 ({\bf r})
\;,
\end{eqnarray}
with
\begin{equation}   \label{eq:Apm}
A_{\pm} (S) = {1\over 2} \left[ \frac{1}{\sqrt{1+S}} \pm
\frac{1}{\sqrt{1-S}} \right] \;. 
\end{equation}
In order to calculate the parameters one has to evaluate integrals
involving Gaussian functions, which have been discussed in Ref.\
\onlinecite{goleb}. If the ion charge is $Z|e|$, and defining
$\beta=\hbar^2/2m$, $\eta=-Ze^2$, and $F_0 (x) \equiv
x^{-1/2}{\rm Erf}(x^{1/2})$, we obtain
\begin{eqnarray} \label{eq:params} 
\epsilon_0 &=&
\beta \left[3 \Gamma^2 + {{a^2 \Gamma^4 S^2} \over { (1-S^2)}} \right]
+ \eta \left({{2 \Gamma \sqrt{2/\pi} } \over{1-S^2}} \right) \left[ 1
+ F_0(2a^2\Gamma^2)- 2 S^2 F_0 (a^2 \Gamma^2 /2) \right] \;,\\
 t & = &
\beta\left[{{a^2 \Gamma^4 S} \over {(1-S^2)}} \right] -\eta\left( {{ 2
\Gamma \sqrt{2/\pi} }\over {1-S^2}} \right) \left[ 2 S F_0 (a^2
\Gamma^2 / 2) - S - S F_0 (2 a^2 \Gamma^2) \right] \;,\\
 X & = &
- e^2\left[ { \Gamma/\sqrt{\pi} S \over (1-S^2)^2} \right]
\left[1+2S^2 + F_0(\Gamma^2 a^2) - 2(1+S^2) F_0(\Gamma^2 a^2/4)
\right] \;,\\
U & = &
e^2 \left[{{ \Gamma/\sqrt{\pi} } \over {(1-S^2)^2}} \right]
\left[ 2 - S^2 + 2S^4 +
 S^2 F_0(\Gamma^2 a^2) - 4S^2 F_0(\Gamma^2 a^2/4) \right] \;,\\
V & = &
e^2\left[  { \Gamma/\sqrt{\pi} \over (1-S^2)^2} \right]
 \left[ S^2 (1+2S^2) + 
(2-S^2) F_0(\Gamma^2 a^2) - 4S^2 F_0(\Gamma^2 a^2/4) \right] \;,\\
 J_z &=& J_{xy} = P =
e^2\left[ { \Gamma/\sqrt{\pi} S^2 \over (1-S^2)^2}\right]
 \left[ 3 + F_0(\Gamma^2 a^2) - 4 F_0(\Gamma^2 a^2/4) \right] \;.
\end{eqnarray}

The parts of the Hamiltonian involving phonons are a free-phonon term
$H^0_{\text{ph}}$ and a Holstein-type interaction $H_{\text{ep}}$, reading
\begin{equation} \label{eq:hph0}
H^0_{\text{ph}}= {{1} \over{2M}}\bigl(P_1^2 +P_2^2 \bigr)+
{{M\Omega^2} \over{2}}\bigl(u_1^2 +u_2^2 \bigr)
\end{equation}
and
\begin{equation} \label{eq:hep}
H_{\text{ep}}\equiv g \sum_\sigma (n_{1\sigma}u_1 + n_{2\sigma} u_2) \;,
\end{equation}
where $M$ is the mass of the ions, $\Omega$ is the phonon frequency,
assumed dispersionless for simplicity (Einstein phonon), and
$P_i \equiv -i \hbar \partial/\partial u_i$ is the
momentum of the ion at position $i$. We are considering only
longitudinal vibrations of the dimer.

The standard procedure to quantize the phononic Hamiltonian  requires 
the introduction of usual phonon operators 
\begin{equation}
\label{eq:phonop_a}
a_i \equiv (2M\hbar\Omega)^{-1/2} \left( M\Omega u_i + i P_i \right)
\end{equation}
for each site. Actually, it is more convenient to work with
symmetric ($s$) and antisymmetric ($b$) combinations
\begin{eqnarray} \label{eq:phonops_sb}
s &=& (a_1 + a_2)/\sqrt{2} \;,\nonumber \\
b &=& (a_1 - a_2)/\sqrt{2} \;,
\end{eqnarray}
such that the free-phonon Hamiltonian $H^0_{\text{ph}}$ can be decomposed as
\begin{equation} \label{eq:freeHph}
 H^{0}_{\text{ph}} \equiv H^{0b}_{\text{ph}}+H^{0s}_{\text{ph}}
= \hbar\Omega  \left( b^\dagger b^{}+{1 \over 2}\right)
+\hbar\Omega  \left( s^\dagger s^{}+{1 \over 2}\right) \;,
\end{equation}
and the electron-phonon Hamiltonian $H_{\text{ep}}$ has two contributions, 
\begin{eqnarray} \label{eq:hep_sb}
H^{s}_{\text{ep}} &=& \hbar\Omega \gamma_0 (n_1 + n_2) \, (s^\dagger +
s^{}) \;, \nonumber \\
H^{b}_{\text{ep}} &=& \hbar\Omega\gamma_0 (n_1 - n_2) \, (b^{\dagger} +b^{}) 
\;,
\end{eqnarray}
where we have used the notation $n_i \equiv \sum_\sigma n_{i\sigma}$,
and introduced the coupling energy per unit charge $g_0\equiv g
\sqrt{\hbar/2M\Omega}$ which, conveniently scaled by the phonon
energy, yields the dimensionless quantity $\gamma_0\equiv g_0/\hbar
\Omega$.  Eqs.\ (\ref{eq:hep_sb}) show explicitly the coupling of $s$
with the total charge, and $b$ with the charge transfer.

\section{The effective electronic Hamiltonian}
\label{sec:Heff}

Following the procedure of Ref.\ \onlinecite{zh91}, we will obtain the
 effective electronic (polaronic) Hamiltonian $H_{\text{pol}}$ by
 first performing on $H$ a generalized Lang-Firsov
 (displaced-oscillator) transformation.  Then, in the transformed
 Hamiltonian, we will take the average of the phononic parts over the
 squeezed-phonon wave function to eliminate the phonon degrees of
 freedom.

\medskip

 As we have two different boson operators, $s$ and $b$, there will be
two corresponding generators $R_s$ and $R_b$, and we have to perform a
sequence of two unitary transformations on $H$.   The generators are
defined as
\begin{equation} \label{eq:Rb}
R_b = \delta\gamma_0
 (n_1-n_2)\, (b^\dagger-b^{})
\end{equation}
and
\begin{equation} \label{eq:Rs}
R_s = \eta\gamma_0
(n_1+n_2)\, (s^\dagger-s^{}) \;, 
\end{equation}
where $\delta$ and $\eta$ are so far undetermined parameters, whose
value will be fixed later in a convenient way.

 Notice that $[R_b,R_s]=0$. Then, the transformed phonon operators are
\begin{equation} \label{eq:btransf}
e^{R_b} b e^{-R_b} =b-\delta \gamma_0 (n_1-n_2)
\end{equation}
and
\begin{equation} \label{eq:stransf}
e^{R_s} s e^{-R_s} =s-\eta \gamma_0 (n_1+n_2) \;.
\end{equation}
We transform, therefore, $H^{0b}_{\text{ph}}+H_{\text{ep}}^b$ as
\begin{eqnarray} \label{eq:hepbtransf}
e^{R_b}(H_{\text{ph}}^{0b}+H_{\text{ep}}^{b})e^{-R_b} &=& \hbar\Omega
\left(b^\dagger b+{1 \over 2}\right)+
\hbar\Omega(1- \delta)\gamma_0 (n_1 - n_2) (b^\dagger +b)
 \nonumber\\ && \mbox{~} 
-\hbar \Omega \delta (2-\delta) \gamma_0^2 (n_1 -n_2)^2
\end{eqnarray}
and
\begin{eqnarray} \label{eq:hepstransf}
e^{R_s} (H^{0s}_{\text{ph}} + H^{s}_{\text{ep}})e^{-R_s} &=& \hbar\Omega
\left(s^\dagger s+{1 \over 2}\right)+
\hbar \Omega (1 - \eta)\gamma_0 (n_1 + n_2) (s^\dagger +s) 
\nonumber\\ && \mbox{~~} 
- \hbar \Omega \eta(2-\eta) \gamma_0^2 (n_1 + n_2)^2 \;.
\end{eqnarray}

Since the purely electronic terms of the total Hamiltonian commute
with the total number of electrons, they are not affected by
$\exp(R_s)$, and are, thus, insensitive to the value of $\eta$. We
then choose $\eta = 1$ in order to eliminate the interaction of
electrons with symmetrical phonons from the transformed Hamiltonian
(see Eq.\ \ref{eq:hepstransf}). We can not do the same with $\delta$,
since $R_b$ does affect the purely electronic terms. Our approach
consists in treating $\delta$ as a variational parameter, to be
determined by minimizing the total energy.

The $R_b$-transformed Fermi operators read
\begin{eqnarray} \label{eq:ctransf}
e^{R_b} c^{}_{1\sigma} e^{-R_b} &=&  c^{}_{1\sigma} \left[ {\rm Ch}
(\gamma_0 B) - {\rm Sh} (\gamma_0 B) \right] \;, \nonumber \\
e^{R_b} c^{}_{2\sigma} e^{-R_b} &=&  c^{}_{2\sigma} \left[ {\rm Ch}
(\gamma_0 B) + {\rm Sh} (\gamma_0 B) \right] \;,
\end{eqnarray}
where we have defined the anti-Hermitian operator $B\equiv
\delta(b^\dagger-b)$, and ${\rm Sh}(x),{\rm Ch}(x)$ are the hyperbolic
sine and cosine functions. Notice the asymmetry with respect to site
exchange due to the fact that the antisymmetric phonon operator has
been defined as a difference of local operators in a well defined
order (see Eq.\ \ref{eq:phonops_sb}).

Using Eqs.\ (\ref{eq:ctransf}), we can easily work out the
transformation of $H_{\text{el}}$ under $R_b$.  Notice that electron number
operators remain unchanged, and the only non-invariant terms of
$H_{\text{el}}$ are 
\begin{eqnarray}
e^{R_b} (c_{1\sigma}^\dagger c_{1\sigma}^{} + H. c.) e^{-R_b} 
&=& {\rm Ch} (2\gamma_0 B) 
(c_{1\sigma}^\dagger c_{1\sigma}^{} + H. c.) + \ldots 
\end{eqnarray}
and
\begin{eqnarray}
e^{R_b} (c_{1\uparrow}^\dagger c_{1\downarrow}^\dagger 
c_{2\downarrow}^{} c_{2 \uparrow}^{} + H.c.) e^{-R_b} &=&
{\rm Ch} (4\gamma_0 B) (c_{1\uparrow}^\dagger c_{1\downarrow}^\dagger
c_{2 \downarrow}^{} c_{2 \uparrow}^{}  + H.c.) + \ldots \;,
\end{eqnarray}
where the dots stand for terms that contain ${\rm Sh} (\mbox{const.}
\times B)$, which average to zero in the squeezed-phonon state, as
discussed below.

\medskip

To eliminate the phonon operators, we take the average of the
transformed Hamiltonian over the squeezed-phonon wave function
\begin{eqnarray}
|\Psi_{\text{sq}} \rangle = e^{-\alpha (bb - b^\dagger b^\dagger)}
|0\rangle_{\text{ph}} \;.
\end{eqnarray}
It is then straightforward to show that
\begin{eqnarray}
\langle \Psi_{\text{sq}}| {\rm Sh}(n \gamma_0 B) | \Psi_{\text{sq}}
\rangle &=& 0 \;, \nonumber \\
\langle \Psi_{\text{sq}}| {\rm Ch}(n \gamma_0 B) | \Psi_{\text{sq}}
\rangle &=& \exp\left[- \frac{1}{2} n^2 \delta^2 \gamma_0^2
e^{-4\alpha} \right] \;.
\label{eq:sqChSh}
\end{eqnarray}
 The result is a polaronic effective Hamiltonian $ H_{\text{pol}} =
H_{\text{el}} (\epsilon_0^\ast, t^\ast, X^\ast, U^\ast, V^\ast,
J^\ast_{xy}, J^\ast_z, P^\ast)+ E_{\text{ph}}$, where
\begin{eqnarray} \label{eq:newparam}
&& \epsilon^\ast_0 = \epsilon - \hbar\Omega \gamma_0^2
[1+\delta(2-\delta)] \;, \quad 
t^\ast=\tau t \;, \quad X^\ast =\tau X \;, \quad
U^\ast = U - 2 \hbar \Omega \gamma_0^2 [1+\delta(2-\delta)] \;,
\nonumber \\ && \mbox{\quad\quad}
 V^\ast = V - 2 \hbar \Omega \gamma_0^2 [1-\delta(2-\delta)] \;,
\quad  P^\ast = \tau^4 P \;, \quad E_{\text{ph}} = \hbar
\Omega \left [{\rm Sh}^2 (2 \alpha) +1 \right] \;,  
\end{eqnarray}
with $\tau \equiv \exp\left[-2 \delta^2 \gamma_0^2 e^{-4\alpha}
\right]$. The parameters $J_z$ and $J_{xy}$ remain unchanged (thus,
they are no longer equal to $P^\ast$). Notice that the energy of the
phonons in the ground-state is given by the last term in Eq.\
\ref{eq:newparam}. Leaving aside $\hbar\Omega/2$, coming from the
$s$-phonons, the contribution of the $b$-phonons to this energy can be
written as
\begin{equation} \label{eq:Ephb}
E_{\text{ph}}^b = \frac{1}{2} \hbar\Omega \, {\rm 
Ch}(4\alpha)
\;.
\end{equation}
Thus,  due to the squeezing
effect, we find an increase of the zero-point $b$-phonon energy, as if the 
frequency of the latter were  
$\Omega^\ast \equiv \Omega \, {\rm Ch} (4\alpha)$.
 This renormalization effect is not directly due to the electronic
interactions, since it does not depend explicitly on them. There
is, however, an indirect influence due to $\alpha$ being
 determined by minimization
of the total energy, which includes the electronic 
part.\cite{omegaeff}

The polaronic Hamiltonian is easily diagonalized, and its energy
eigenvalues and eigenvectors are listed in Table \ref{tab:eigen}.
Leaving aside differences in notation, these results are consistent
with those obtained by other authors.\cite{js81,deboer}.

\section{Correlation functions}
\label{sec:cf}

The quantities accessible to measurement, besides the energy
eigenvalues, are expressed in the form of various correlation
functions (CF), which can be evaluated through the eigenstates of the
effective Hamiltonian, listed in Table \ref{tab:eigen}.  We will
indicate by $\langle\langle X \rangle\rangle_{|\ell\rangle}$ the CF
for a generic operator $X$ in one of the eigenstates $|\ell\rangle$ of
the effective Hamiltonian.  The double average $\langle\langle X
\rangle \rangle_{|\ell\rangle}$ has to be understood as follows:
\begin{equation}
\langle \langle X \rangle \rangle_{|\ell\rangle} \equiv 
\langle \ell|\langle \Psi_{\text{sq}} | e^{R_b} e^{R_s} 
X e^{-R_s} e^{-R_b} |\Psi_{\text{sq}}\rangle |\ell\rangle \;.
\end{equation}
In particular, $\langle \langle H \rangle \rangle_{\vert \ell\rangle}
= E^\ast_\ell$.  We will evaluate a number of correlation functions
which have been considered in the literature, whose definitions are 
 briefly clarified below. \par

The squeezing parameter is directly accessible to experimental measure
through the Debye-Waller factor, defined as
$F^{DW}\equiv\langle\langle u_i^2\rangle\rangle-\langle\langle
u_i\rangle\rangle^2$. Indeed, by using the time-dependent
representation of the displacement through the effective Hamiltonian
$H^\ast$, i.e.  $u_i(t)\equiv \exp(iH^\ast t/\hbar) u_i\exp(-iH^\ast
t/\hbar)$ one obtains
 \begin{equation} F^{DW}={{L^2}\over
{2}}\left[1+e^{4\alpha}\right] \label{DebyeWaller}
\end{equation} 
where $L\equiv \sqrt{\hbar/2M\Omega}$ is the characteristic phonon
length.

An important quantity is the ratio of the kinetic energy of the
interacting system to that on the non-interacting one, defined as
\begin{equation}    
F^{\text{kin}}_{|\ell\rangle}\equiv
{{\langle \langle  t\sum_\sigma ( a^\dagger_{1\sigma}a^{}_{2\sigma}
+a^{\dagger}_{2\sigma} a^{}_{1\sigma})\rangle\rangle_{\gamma_0\ne 0}}
\over{[\langle \ell|t\sum_\sigma( a^\dagger_{1\sigma}a^{}_{2\sigma}
+a^{\dagger}_{2\sigma} a^{}_{1\sigma})|\ell\rangle]_{\gamma_0=0} }}
 \label{cfkin1}
\end{equation}
 For $N=1$, $F^{\text{kin}}_{|\ell\rangle}$ is just the hopping reduction
 factor $\tau$.  For $N=2$, $F^{\text{kin}}_{|\ell\rangle}$ vanishes, except
 in the state $|Sb\rangle$, for which, defining $\theta^\ast$ like
 $\theta$ in Table \ref{tab:eigen}, but using the renormalized
 interactions, we obtain
\begin{equation}
F^{\text{kin}}_{|Sb\rangle}={{\tau\sin(2\theta^\ast)} 
\over{\sin(2\theta)}} \;.
\label{cftkin4}
\end{equation}
 
 To measure the relation between the charge and the phonon
number on each site (related to the polarizability of the medium)
Ref.\ \onlinecite{wellein} introduced, for the case of one electron,
the ``electron-phonon CF''.  In our case it has to be defined as
\begin{equation} 
F^{\text{ep}}_{jk} \equiv
\langle \langle  n_j b^\dagger_kb^{}_k \rangle\rangle 
\,, \qquad (j,k=1,2)\,.
\label{eq:Felphdef}
\end{equation}
As $k$ can coincide or not with $j$, we will consider both on-site
($j=k$) and inter-site $(j\ne k)$ electron-phonon CF's.  The
inter-site charge CF is defined as $\frac{1}{4} \langle \langle n_1
n_2 \rangle \rangle$, and the longitudinal and transverse spin CF's
are $\langle\langle S^z_1 S^z_2 \rangle\rangle$ and $\langle\langle
S^-_1 S^+_2 \rangle\rangle$, respectively.  The {\it bipolaron\/}
``mobility'', or {\it singlet-superconductor\/} CF,\cite{luz})
describes the hopping of two particles occupying the same site,
namely,
\begin{eqnarray}  \label{eq:osbip1}
\langle\langle  (c_{2\downarrow}^{} c_{2 \uparrow}^{}
c_{1\uparrow}^\dagger c_{1\downarrow}^\dagger)
 \rangle\rangle &=&
\langle \Psi_{\text{sq}}|{\rm Ch}[4\gamma_0 \delta(b^\dagger-b)]
|\Psi_{\text{sq}}\rangle 
\langle\ell | c_{2\downarrow}^{} c_{2 \uparrow}^{}
c_{1\uparrow}^\dagger c_{1\downarrow}^\dagger |\ell\rangle 
\nonumber \\
 &=& \tau^4 \langle\ell | c_{2\downarrow}^{} c_{2 \uparrow}^{}
c_{1\uparrow}^\dagger c_{1\downarrow}^\dagger |\ell\rangle \;.
\end{eqnarray}
Finally, the (phonon-induced) average charge transfer is defined,
following Ref.\ \onlinecite{zh91}, as
\begin{equation}
\label{eq:fctdef}
F^{\text{CT}}_{|\ell\rangle} \equiv {{\langle \langle (n_1 - n_2)
(b^\dagger + b) \rangle \rangle_{|\ell\rangle} } \over {\sqrt{\langle
\langle (b^\dagger + b)^2 \rangle\rangle_{|\ell\rangle} } }} = {{-2
\delta\gamma_0 \langle\ell|(n_1 -n_2)^2 |\ell\rangle } \over{ \sqrt{
{\rm Ch}(4\alpha) +4 \delta^2 \gamma_0^2 \langle \ell |(n_1 - n_2)^2
|\ell\rangle} }} \;.
\end{equation}
The analytical expressions of various CF's are collected in Table
\ref{tab:correl}.

\section{Results}       
\label {sec:results}
The ground state of $H_{\text{pol}}$ for several values of $g_0$ was
identified by searching the minimum of the total, i.e. electronic plus
phononic, energy upon independent and simultaneous optimization of the
parameters that define the shape of the orbitals ($\Gamma$), the
displaced-oscillator strength ($\delta$), and the squeezing-effect
strength ($\alpha$).  Given $\alpha$ and $\delta$, then the optimal hopping 
reduction factor $\tau\equiv \exp(-2\delta^2\gamma_0^2e^{-4\alpha})$ 
follows.  Specifically, the optimization proceeded in
three steps. First, the total energy for each eigenvalue was
separately optimized with respect to $\Gamma,\delta$ and
$\alpha$. Second, the minimal total energy was selected with the
corresponding values of the variational parameters. Third, the
higher-lying eigenvalues where recalculated with $\Gamma,\delta$ and
$\alpha$ set by the optimization of the lowest-lying one. This
procedure gave us the true full spectrum of eigenvalues for each
ground state.\par
  We will now discuss the results for $N=1$ (quarter
filled case) and $N=2$ (half filled case).

\subsection{Quarter filled case}  \label{subsec:N1}

The interplay between the electronic and phononic subsystems is 
evidenced in  Fig.\ \ref{fig:Gam_del_alph_N1} which shows $a\Gamma,\delta$ 
and $\exp(-4\alpha)$ as functions of $a$, for $g_0=0.447\,$eV and
$\hbar\Omega =0.1\,$eV. The displacement parameter $\delta$  
grows smoothly with $a$, until, at a critical length
$a_{c}\approx 2.1$\AA, it jumps discontinuously up to $\delta\simeq 1$. 
The squeezing-related quantity $\exp(-4\alpha)$ decreases with $a$ up 
to $a_{c}$, indicating that the squeezing effect becomes stronger to 
counteract the growing displacement.  At $a=a_{c}$, $\exp(-4\alpha)$ 
recovers abruptly a rather large value. The orbital-shape
parameter $a\Gamma$ rises almost linearly with $a$, except around 
$a_{c}$ where it suffers a small, sharp increase.  Therefore, for $N=1$, 
the changes of the phononic parameters have only a small effect on 
$\Gamma$. 

The changes in the electron and phonon subsystems depicted in Fig.\
\ref{fig:Gam_del_alph_N1} are mirrored by the correlation functions 
shown in Fig.\ \ref{fig:ke_epcf_ct_N1}. We notice that the discontinuities 
in $\delta$ and $\exp(-4\alpha)$ at $a_c$ correspond to the localization 
of the polaron, as signalled by the fact that for $a=a_c$ the kinetic 
energy vanishes (discontinuously) and the on-site electron-phonon 
correlation jumps to a very high value. The localization is driven by 
$\tau = \exp{\left[ -2 \delta^2 \gamma_0^2 e^{-4\alpha} \right] }$, which
renormalizes the effective hopping (see Eq.\~(\ref{eq:newparam})). The
sudden changes observed in Fig.1 can be attributed to the competition 
between the squeezing effect, that tends to keep $\tau$ finite through 
the factor $e^{-4\alpha}$, and the polaron effect, that tends to reduce
$\tau$ through the factor $\delta^2$ in the exponent of $\tau$.
As $\delta$ increases with $a$ (Fig.\ \ref{fig:Gam_del_alph_N1}), 
$\alpha$ also increases to gain electronic energy by keeping $t$ at a 
non-vanishing value (see Table~\ref{tab:eigen}). The price to pay is an 
increase of the phonon energy, Eq.\~(\ref{eq:Ephb}). The behavior shown   
for $a\le a_c$ by all quantities in Fig.2 clearly reflects the continuous 
variations of the phononic parameters of Fig.1. When the phonon energy 
becomes too large, at $a=a_c$, the system jumps to a state with a small 
$\alpha$ and zero hopping. However, if the {\it bare} hopping amplitude 
is by itself small (what happens for large values of $a$), the loss in 
electronic energy when $t$ is reduced is not very important, $\alpha$ 
remains small, and there is no sudden change of state. This can be seen 
in Fig.\ \ref{fig:CFs_N1}, where the kinetic energy and the electron-phonon 
correlation functions are plotted as functions of $g_0$ for two values 
of $a$, the large-$a$ case showing a continuous transition.

The localization critical length depends on the strength of the 
electron-phonon interaction. We can then draw a ``phase-diagram'' for 
the system, as shown in Fig.\ \ref{fig:phasediag_N1}. In both panels, 
the region to the left (right) of each line corresponds to itinerant 
(localized) states. On the left panel we have chosen to plot the
bare hopping parameter $t$, calculated at $a_c$, as a function of
$g_0$. We also show, on the right, that $g_0^2/\hbar\Omega$ is a good
scaling variable for not too large dimer lengths, as the curves for different
phonon frequencies coincide. For large $a$, the curves tend to separate, 
enhancing the curvature so as to avoid crossing the $x$-axis. 
This implies that the polaronic transition as $a$ increases   
becomes a continuous crossover for small values of $g_0$.

\subsection{Half-filled case}  \label{subsec:N2}

The half-filled case presents a more complex behavior than the
quarter-filled one, due to the electron-electron interactions that
become effective in this case. We first discuss the ``phase diagram'',
Fig.\ \ref{fig:phasediag_N2}, in order to give an overall view of the
physics. Again there are two dominant regimes, localized and
non-localized, respectively above and below the continuous line in
Fig.\ \ref{fig:phasediag_N2}. The physics of the system is dominated
by the singlet bonding state $|Sb\rangle$ (see Table~\ref{tab:eigen}),
which is always the lowest-lying energy level. However, for large $a$
$|Sb\rangle$ becomes degenerate with either the triplet states
$|T\rangle$ (Fig.\ \ref{fig:eigen_N2}, top panel) or the charge
transfer states $|CT\rangle$ (Fig.\ \ref{fig:eigen_N2}, middle and
bottom panels), depending on the value of $g_0$. These degeneracies
just mean that these states are no longer appropriate to describe the
system, which has become a pair of isolated atoms. For small $g_0^{}$,
as the dimer lenght increases, the system evolves continuously from an
extended singlet state to a localized state with one electron per
site, that occurs for very large values of $a$. For large $g_0$ a
bipolaron is formed (in the $|CT\rangle$ state) which eventually
localizes in one of the sites, leaving the other one empty.

While one can draw an $\Omega$-independent phase diagram, the
correlation functions do depend on $\Omega$. We have therefore
selected $\hbar\Omega=0.1\,$eV, which might be realistic in HTSC's and
colossal magnetoresistance materials. We will present the results for
$g_0^2/\hbar\Omega = 2.0$, 2.2 and 2.5\,eV, corresponding to
$g_0=0.447$, 0.470 and 0.500\,eV, respectively. The dotted horizontal
lines in Fig.\ \ref{fig:phasediag_N2} visualize the sections of the
phase diagram that we are going to discuss in detail in Fig.\
\ref{fig:eigen_N2}, where the full eigenvalue spectrum is shown as a
function of $a$ for the three mentioned values of $g_0$. When
$g_0^2/\hbar\Omega=2.0\,$eV, $|Sb\rangle$ is always the ground state,
with the $|T\rangle$ state approaching it asymptotically for large $a$
(top panel of Fig.\ \ref{fig:eigen_N2}). For
$g_0^2/\hbar\Omega=2.2\,$eV we find a re-entrant behavior in a range
of values of $a$ in which a rearrangement of the full spectrum
occurs. Finally, in the case $g_0^2/\hbar\Omega=2.5\,$eV one crosses
the region of the phase diagram in Fig.\ \ref{fig:phasediag_N2}
bounded by the broken line, where, as we will see below, both the
(antiferro)magnetic CF and the charge transfer have large values in
the $|Sb\rangle$ state.  Correspondingly, one sees from the bottom
panel of Fig.\ \ref{fig:eigen_N2} that inside this CT-AF region the
state $|CT\rangle$ becomes lower in energy than $|T\rangle$.

The changes in the electronic and phononic parameters underlying the level 
crossing in Fig.\ \ref{fig:eigen_N2} are shown in 
Fig.\ \ref{fig:Gam_del_alph_N2}. In the top panel $a\Gamma$ shows an 
approximately linear trend, steeper for $g_0^2/\hbar\Omega=2.0\,$eV 
than for $g_0^2/\hbar\Omega=2.5\,$eV. This can be explained
by anticipating that, in the latter case, even when, for $a<1.60$\AA, the
stable state is $|Sb\rangle$, still there is an appreciable charge transfer,
indicating that the electrons tend to keep apart. Their charge-density
clouds can, therefore, spread out more largely around the ions (i.e. 
$\Gamma$ can decrease) without too much increase in the Coulomb energy.  
In the intermediate case $g_0^2/\hbar\Omega=2.2$\,eV, $a\Gamma$ jumps between
the two lines. The jumps mark the re-entrant transition between the 
localized and the non-localized regime, which therefore implies marked 
modifications in the Wannier functions shape. The middle panel shows the 
corresponding behavior of the phonon displacement parameter $\delta$. We see 
that the passage from the non-localized regime to the localized one
is accompanied, as expected, by an abrupt jump from $\delta\simeq 0$  
to $\delta\simeq 1$. Only for $g_0^2/\hbar\Omega=2.5\,$eV the limiting 
value $\delta =1$ is reached by two jumps connected by a range of gradual 
increase, associated with the crossing of the CT-AF region in 
Fig.\ \ref{fig:phasediag_N2}. The behavior of the squeezing-related 
parameter $\exp(-4\alpha)$, shown in the bottom panel, is also characterized 
by sudden transitions at the line boundaries. Its dependence on $a$, 
combined with the negligible value that $\delta$ takes when $|Sb\rangle$ 
is the non-degenerate ground state, implies the existence of a threshold 
value for $g_0^2/\hbar\Omega$ below which the electrons influence the 
phonons, but the converse is not true, in the sense that the latter are 
completely screened out, even though $g_0$ is not small. Large $\delta$ 
values in the $|Sb\rangle$ state can only be found in the CT-AF region, 
which, however, is located above the threshold.  

In Fig.\ \ref{fig:params_N2}
we show the three largest effective electronic interactions $t^\ast$, 
$U^\ast$ and $V^\ast$ (top, middle and bottom panel, respectively), for
the chosen values of $g_0$. Besides showing sharp discontinuities at the 
critical values of $a$, their behavior clarifies the nature of the
CT-AF region of the phase diagram. Indeed, we can see that this region
is characterized by a significant reduction of the on-site effective
Coulomb repulsion and an increase of the inter-site one, which are 
necessary conditions for the stabilization of a local bipolaron,
while the effective hopping is only slightly reduced. Concomitantly,
the bipolaron mobility remains finite (Fig.\ \ref{fig:CFs_N2}, bottom 
panel) while charge-transfer fluctuations build up (Fig.\ \ref{fig:CFs2_N2}, 
top panel), and the inter-site spin (or magnetic) correlations are still 
important (Fig.\ \ref{fig:CFs2_N2}, bottom panel). We would like to remark 
that the existence of this region of the phase diagram (probably the most
interesting one as far as high-$T_c$ superconductors are concerned)
depends strongly on the squeezing of the phonon states, as can be seen
in the bottom panel of Fig.\ \ref{fig:Gam_del_alph_N2}. In other
words, the squeezing effect counteracts the oscillator displacement,
and succeeds in preserving a rather itinerant character even in the
presence of both appreciable charge transfer and magnetic
correlations.

\section{Conclusions}
\label{sec:concl}

We have presented here a variational study of a dimer where interacting
electrons occupy a single orbital per site and are coupled to Holstein
phonons. We have put into evidence
that the variation of the dimer length and/or the strength of the
electron-phonon coupling yield a strong renormalization of the
interaction constants as well as the width of the Wannier functions
describing the local orbitals, which establishes a link between the
electronic interactions and the phonons at a deeper level than
predicted by the standard polaron approach. The squeezing of the
phonon states has been shown to be particularly relevant, yielding
sharp changes in the state of the system as the dimer length is varied
at fixed electron-phonon coupling. The experimental interest of this
finding is that situations where the interatomic distances can be
varied, even by a small amount, either uniformly by an external
pressure, or inhomogeneously by dimerization in a chain compound, can
lead to important changes in the system's behavior.

Obviously one has to be careful in making predictions about the
behavior of an extended system based on calculations for a two-site
cluster. Nevertheless, despite the fact that Figs.\
\ref{fig:phasediag_N1} and \ref{fig:phasediag_N2} cannot be viewed as
{\it phase diagrams\/}, we believe that they allow us to speculate on
the nature of the corresponding states in a lattice. Remaining in the
half-filled case, the region where the extended bonding singlet is
stable should show the antiferromagnetic order characteristic of the
usual Hubbard model, with a continuous evolution towards a
local-moment paramagnet as the interatomic distance is increased. For
a doped system, one could expect the two-site singlets to build up a
RVB-like state, or an inter-site bipolaron, in the lattice. On the
other hand, the $|CT\rangle$ state should correspond to a charge
density wave in the lattice. In connection to the HTSC materials, our
most significant finding is that there is a region in the parameter
space where the kinetic energy, the charge transfer and the magnetic
correlations are simultaneously large, while the phonon state is
partially displaced and strongly squeezed. This result suggests that
it is perhaps not necessary to set a sharp alternative between either
magnetic or charge instabilities. Based on our results, one might
speculate that, in the normal state, those materials might be in a
situation corresponding to the one in the dimer, where both types of
fluctuations can be simultaneously large. It is interesting to notice
that the existence of this narrow region (between the dashed and
continuous lines in Fig.\ \ref{fig:phasediag_N2}) in our model is
entirely due to the squeezing of the phonon states. Indeed, we checked
that it disappears if one keeps the squeezing-parameter $\alpha$ fixed
at zero, giving rise to a direct transition from the extended singlet
ground-state ($\delta\approx 0.$) to the charge transfer one
($\delta=1.$). We thus suggest that the competition between magnetic
and charge instabilities in HTSC materials might be driven by
electron-phonon interactions in the presence of strongly squeezed
phonon states. It would be interesting to use the dimer solutions
obtained here as building blocks for lattice states, with the aim of
obtaining quantitative support for our qualitative analysis of the
lattice case. Work on this line is now in progress.

\acknowledgements

We would like to thank J. Spa\l ek, W. W\'ojcik, A. Painelli,
S. Ciuchi, and H. Zheng for many stimulating discussions.  The present
work was financially supported under a CNR(Italy)-CNPq(Brazil) scientific
agreement, grant no.\ 910119/93-7, and by the brazilian agency
CAPES-MEC. M. A. was also partially supported by Istituto Nazionale
per la Fisica della Materia (INFM), Italy, and CAPES-MEC, Brazil.



\begin{table} \squeezetable
\caption{Eigenvalues and eigenvectors of the Hamiltonian of Eq.(1) for
$N=1,2,3,4$. $D$ is the degeneracy of the state. The labels $a,b$
indicate $bonding$ and $antibonding$ character, while $S$ and $T$
correspond to $Singlet$ and $Triplet$ states.  For $N=2$ we have
defined $E_U\equiv 2\epsilon_0+U+P$, $E_V\equiv 2\epsilon_0+V+J_{xy}$,
$r\equiv \sqrt{(E_U-E_V)^2+16(t-X)^2}$, 
$\tan(\theta)\equiv -4(t-X)/(E_U-E_V+r)$ and $T=t-2X$.}
\begin{tabular}{lccl} 
 Filling and Energy &D & $S,\,S_z$ & Eigenvectors \\ \hline \hline 
\rule[-5pt]{0pt}{20pt} ($ N=1$) &&&\\ 
$ E_1=\epsilon_0-t $ &2 & $1/2,1/2$ & $|1b,\uparrow\rangle =
{{1}\over{\sqrt{2}}}[c^\dagger_{1\uparrow}+c^\dagger_{2\uparrow}]
|0\rangle $ \\ && 1/2,-1/2 & $|1b,\downarrow\rangle =
{{1}\over{\sqrt{2}}} [c^\dagger_{1\downarrow} +
c^\dagger_{2\downarrow}]| 0 \rangle $ \\ $ E_2=\epsilon_0+t \,\,\, $
&2 & 1/2,1/2 & $|1a,\uparrow\rangle=
{{1}\over{\sqrt{2}}}[c^\dagger_{1\uparrow}-c^\dagger_{2\uparrow}]
|0\rangle $ \\ & & 1/2,-1/2 & $|1a,\downarrow\rangle=
{{1}\over{\sqrt{2}}}[c^\dagger_{1\downarrow}-c^\dagger_{2\downarrow}]|
0 \rangle $ \\ \\ \hline \rule[-5pt]{0pt}{20pt}  ($ N=2$) & & & \\
$E_{Sb}=\frac{1}{2}(E_U+E_V-r)$ &1 & $0,0$ &
$|Sb{\rangle}=\frac{1}{\sqrt{2}}[\sin \theta (c^\dagger_{1\downarrow}
c^\dagger_{1\uparrow} + c^\dagger_{2\downarrow} c^\dagger_{2\uparrow})
$\\ & & & $ \mbox{~~~~~~~~~} - \cos \theta (c^\dagger_{1\downarrow}
c^\dagger_{2\uparrow} + c^\dagger_{2\downarrow}
c^\dagger_{1\uparrow})]|0{\rangle}$\\ & && \\ $E_{T,\pm
1}=2\epsilon_0+V-J_z$ & 2 & $1,1 (-1)$ & $|T,\pm
1{\rangle}=c^\dagger_{1\uparrow(\downarrow)} c^\dagger_{2\uparrow
(\downarrow)}|0{\rangle}$\\ $E_{T,0}=2\epsilon_0+V-J_{xy}$ & 1 & $1,0$
& $|T,0{\rangle}=\frac{1}{\sqrt{2}}[c^\dagger_{2\downarrow}
c^\dagger_{1\uparrow} - c^\dagger_{1\downarrow}
c^\dagger_{2\uparrow}]|0{\rangle}$\\ & & \\ $E_{CT}=2\epsilon_0+U-P$ &
1 & $0,0$ & $ |CT{\rangle} =
\frac{1}{\sqrt{2}}[c^\dagger_{1\downarrow} c^\dagger_{1\uparrow} -
c^\dagger_{2\downarrow} c^\dagger_{2\uparrow}]|0{\rangle}$\\ & & &\\
$E_{Sa}=\frac{1}{2}(E_U+E_V+r)$ &1 & $0,0$ &
$|Sa{\rangle}=\frac{1}{\sqrt{2}}[\cos \theta (c^\dagger_{1\downarrow}
c^\dagger_{1\uparrow} + c^\dagger_{2\downarrow} c^\dagger_{2\uparrow})
$\\ & & & $ \mbox{~~~~~~~~~} + \sin \theta (c^\dagger_{1\downarrow}
c^\dagger_{2\uparrow} + c^\dagger_{2\downarrow}
c^\dagger_{1\uparrow})]|0{\rangle}$\\ \\ \hline 
 \rule[-5pt]{0pt}{20pt} ($ N=3$) & & & \\ $
E_{3b\sigma}= 3\epsilon_0+U+2V-J_z-T $ & 2 & 1/2,1/2 &
$|3b,\uparrow\rangle= {{1}\over{\sqrt{2}}} [c^\dagger_{1\uparrow}
c^\dagger_{1\downarrow} c^\dagger_{2\uparrow} +c^\dagger_{2\uparrow}
c^\dagger_{2\downarrow} c^\dagger_{1\uparrow} ]|0\rangle$ \\ & &
1/2,-1/2 & $|3b,\downarrow\rangle= {{1}\over{\sqrt{2}}}
[c^\dagger_{1\uparrow} c^\dagger_{1\downarrow} c^\dagger_{2\downarrow}
+c^\dagger_{2\uparrow} c^\dagger_{2\downarrow} c^\dagger_{1\downarrow}
]|0\rangle$ \\ $ E_{3a}=3\epsilon_0+U+2V-J_z+T $ &2 & 1/2,1/2 &
$|3a,\uparrow\rangle= {{1}\over{\sqrt{2}}} [c^\dagger_{1\uparrow}
c^\dagger_{1\downarrow} c^\dagger_{2\uparrow} -c^\dagger_{2\uparrow}
c^\dagger_{2\downarrow} c^\dagger_{1\uparrow} ]|0\rangle$ \\ & &
1/2,-1/2 & $|3a,\downarrow\rangle= {{1}\over{\sqrt{2}}}
[c^\dagger_{1\uparrow} c^\dagger_{1\downarrow} c^\dagger_{2\downarrow}
-c^\dagger_{2\uparrow} c^\dagger_{2\downarrow} c^\dagger_{1\downarrow}
]|0\rangle$ \\ \\ \hline \rule[-5pt]{0pt}{20pt}  ($ N=4$) & & & \\ 
$E_4=4\epsilon_0+2U+4V-2J_z$ &
1 & 0,0 & $|4 \rangle= c^\dagger_{1\uparrow}
c^\dagger_{1\downarrow}c^\dagger_{2\uparrow}
c^\dagger_{2\downarrow}|0\rangle $ \\ & & \\ 
\end{tabular}
\label{tab:eigen} 
\end{table}

\begin{table} \squeezetable
\caption{Analytic expressions of some correlation functions
$\langle\langle X \rangle\rangle$ in the indicated eigenstates.}
\begin{tabular}{c||c|c|c|c|c}
&&&&&\\
X & $|1\sigma\rangle$&
$|Sb\rangle$ & $|T,\pm 1\rangle$ & $|T, 0\rangle$  & $|CT\rangle$ \\
& & && &\\ 
\hline \hline && && \\
$ n_jb^\dagger_jb^{}_j $  
 & $ \frac{1}{4} \bigl[\, \sinh^2(2\alpha) $ ~~~
 & $ \frac{1}{4} \sinh^2(2\alpha) + \gamma_0^2$ ~~~
& $ \frac{1}{4} \sinh^2(2\alpha) $ ~~
& $ \frac{1}{4} \sinh^2(2\alpha) $ ~~
& $ \frac{1}{4} \sinh^2(2\alpha) $ ~~~
\\  
($j=1,2$) 
& $ +\gamma_0^2(1+\delta)^2 \,\bigr] $
 & $+\gamma_0^2\delta(2+\delta)\sin^2\theta$ 
&$+\gamma_0^2$ 
& $+\gamma_0^2$ 
& $+\gamma_0^2(1+\delta)^2 $ \\
&&&&&\\ 
\hline & && && \\
$ n_jb^\dagger_kb^{}_k $  
 & $ \frac{1}{4} \bigl[\, \sinh^2(2\alpha) $ ~~~
 & $ \frac{1}{4} \sinh^2(2\alpha) +\gamma_0^2$ ~~~
& $ \frac{1}{4} \sinh^2(2\alpha) $ ~~
& $ \frac{1}{4} \sinh^2(2\alpha) $ ~~
& $ \frac{1}{4} \sinh^2(2\alpha)$ ~~~
\\  
($j\ne k=1,2$)
&$+\gamma_0^2(1-\delta)^2 \,\bigr] $
 & $-\gamma_0^2\delta(2-\delta)\sin^2\theta$ 
&  $+\gamma_0^2$ 
& $+\gamma_0^2$ 
& $+\gamma_0^2(1-\delta)^2 $ \\
&&&&&\\ 
\hline & && && \\
$ \frac{1}{4} n_1 n_2 $
& 0.
 & $ \frac{1}{4} \cos^2\theta $
 &  $\frac{1}{4}  $
&  $ \frac{1}{4} $
&  0. \\
&&&&&\\ 
\hline & && && \\
 $ S^z_1S^z_2 $
& 0.
 & $ - \frac{1}{4} \cos^2\theta $
& $ \frac{1}{4} $
 & $- \frac{1}{4} \cos^2\theta $ ~~
& 0. \\
&&&&&\\
\hline & && && \\
$ S^-_1 S^+_2 $
& 0.
 & $- \frac{1}{2} \cos^2\theta  $
& $0.$ 
&$ - \frac{1}{2} $
& $0.$ \\ 
&&&&&\\
\hline & && && \\
$c_{2\downarrow}^{}c_{2\uparrow}^{}c_{1\uparrow}^{\dagger}
c_{1\downarrow}^{\dagger}
$
&$0.$
 & $  \frac{1}{2} \tau^4 \sin^2\theta $
& $0.$  
& 0.
& $- \frac{1}{2}\tau^4 $ \\ 
&&&&& \\
\hline 
&&&&&\\
$(n_1-n_2)(b^\dagger-b)$
&$ -2\delta\gamma_0 $
 & $-16\delta\gamma_0\sin^2\theta$
& $0.$  
& $0$.
 & $-16\delta\gamma_0$ \\ 
$\overline{\sqrt{\langle\langle(b^\dagger-b)^2\rangle\rangle}}$
& $\overline{\sqrt{\cosh(4\alpha)+(2\delta\gamma_0)^2}}$
& $\overline{\sqrt{\cosh(4\alpha)+(4\delta\gamma_0\sin\theta)^2 }}$
&&
& $\overline{ \sqrt{\cosh(4\alpha)+(4\delta\gamma_0)^2} }$ \\
&&&&&
\end{tabular} 
\label{tab:correl}
\end{table}

\begin{figure}
\caption{The Wannier-function-shape parameter $a\Gamma$, the
displaced-oscillator parameter $\delta$, and the squeezing parameter
$\exp(-4\alpha)$ as functions of $a$, for $N=1$,
$\hbar\Omega=0.1\,$eV, and $g_0 = 0.447\,$eV.}
\label{fig:Gam_del_alph_N1}
\end{figure}

\begin{figure}
\caption{Variation of the reduced kinetic energy, the charge transfer,
 and the on-site and inter-site EP CF's with $a$, for $N=1$,
 $\hbar\Omega=0.1\,$eV and $g_0=0.447\,$eV.}
\label{fig:ke_epcf_ct_N1}
\end{figure}

\begin{figure}
\caption{The same quantities as in Fig.\
\protect\ref{fig:ke_epcf_ct_N1} as functions of $g_0$, for fixed
length values $a=2.3\,$\AA\ and $a=4.0\,$\AA. The inter-site EP CF for
$a=4.0\,$\AA\ has vanishingly small values on the scale of the
figure.}
\label{fig:CFs_N1}
\end{figure}

\begin{figure}
\caption{The ``phase diagram'' for $N=1$ and two values of
$\hbar\Omega$.  The vertical coordinate is the bare hopping parameter
$t$, and the horizontal one is the electron-phonon coupling $g_0$
(left panel). The electron is itinerant on the left of each curve, and
localized on the right. The right panel shows that $g_0^2/\hbar\Omega$
is a good scaling variable for small $a$, as the two curves
coincide.}
\label{fig:phasediag_N1}
\end{figure}

\begin{figure}
\caption{The ``phase diagram'' for $N=2$. The electrons are
non-localized below the full line, and localized in the region above
it. The broken line delimits the region where both the magnetic CF and
the CT are large. The horizontal lines correspond (for
$\hbar\Omega=0.1$\,eV) to the values of $g_0^2/\hbar\Omega$ considered
in the numerical analysis.}
\label{fig:phasediag_N2}
\end{figure}

\begin{figure}
\caption{Dependence of the full eigenvalue spectrum for $N=2$ with $a$,
 for $\hbar\Omega=0.1$\,eV, and the indicated values of $g_0$.}
\label{fig:eigen_N2}
\end{figure}

\begin{figure}
\caption{Microscopic electronic and phononic parameters for $N=2$ vs.\
$a$ for the indicated values of $g_0$. Top: the Wannier wavefunction
shape parameter $a\Gamma$; middle: the oscillator displacement $\delta$;
bottom: the squeezing parameter $\exp(-4\alpha)$.}
\label{fig:Gam_del_alph_N2}
\end{figure}

\begin{figure}
\caption{The effective hopping parameter $t^\ast$ (top), the on-site
interaction $U^\ast$ (middle), and the inter-site interaction $V^\ast$
(bottom) as functions of $a$, for $N=2$, $\hbar\Omega=0.1$\,eV, and the
indicated values of $g_0$.}
\label{fig:params_N2}
\end{figure}

\begin{figure}
\caption{The on-site electron-phonon CF (top), the inter-site
electron-phonon CF (middle), and the singlet-superconductor CF (bottom) as
functions of $a$, for $N=2$, $\hbar\Omega=0.1$\,eV the indicated values
of $g_0$.}
\label{fig:CFs_N2}
\end{figure}

\begin{figure}
\caption{The charge transfer (top) and the absolute value of the
magnetic CF, $|\langle\langle S^z_1S^z_2 \rangle\rangle|$, (bottom)
vs.\ $a$, for $N=2$, $\hbar\Omega=0.1$\,eV, and the indicated
values of $g_0$.}
\label{fig:CFs2_N2}
\end{figure}

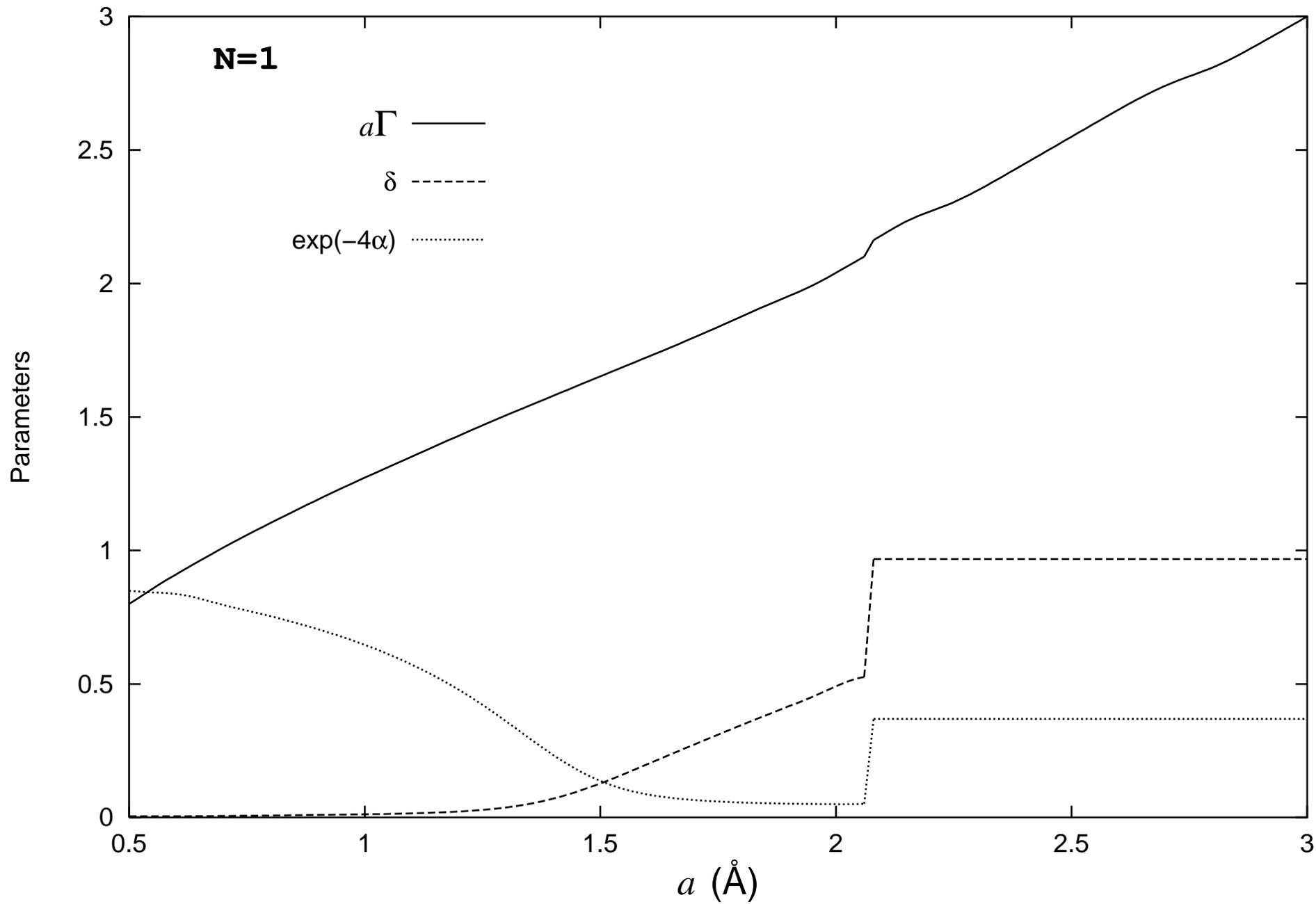
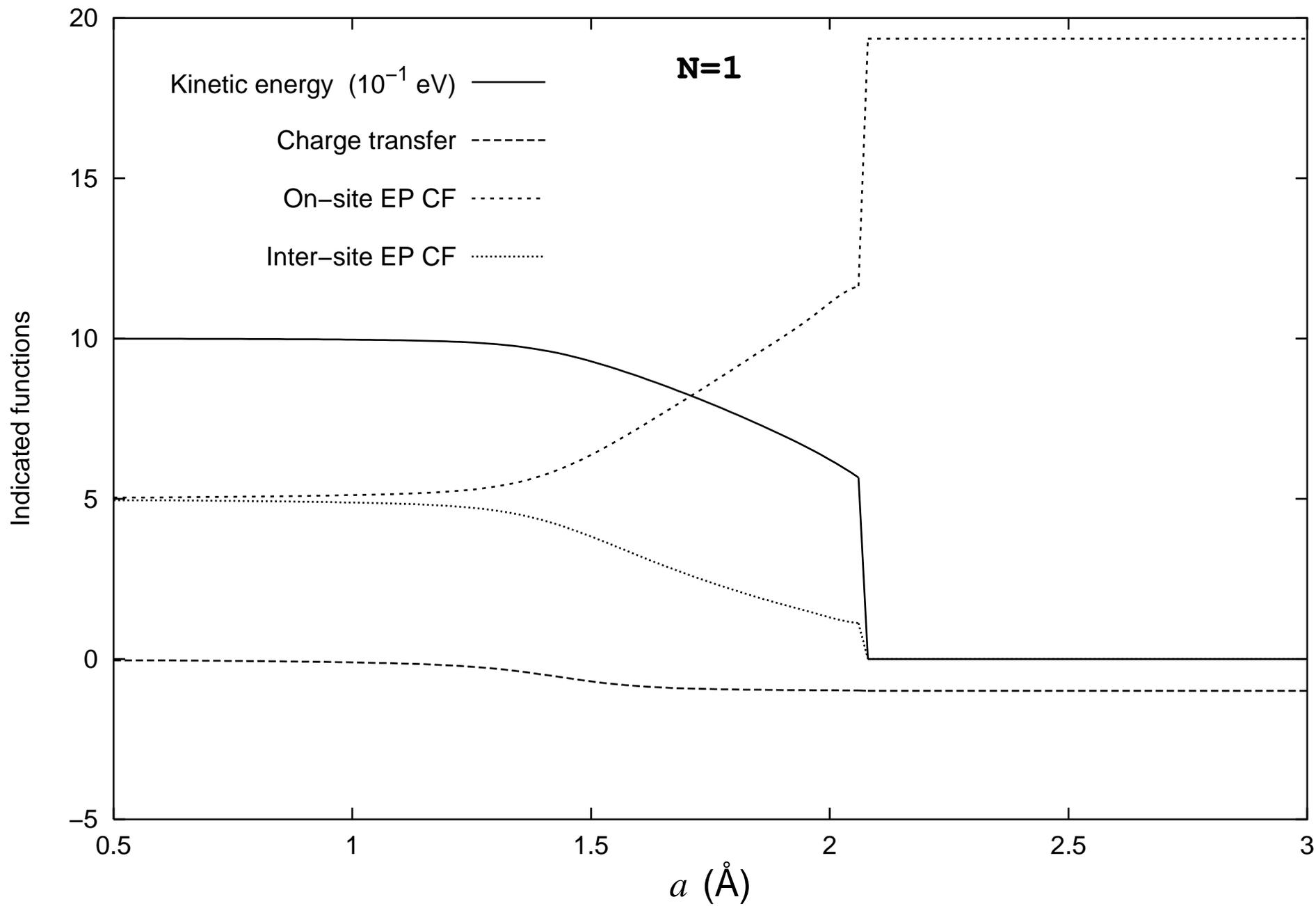
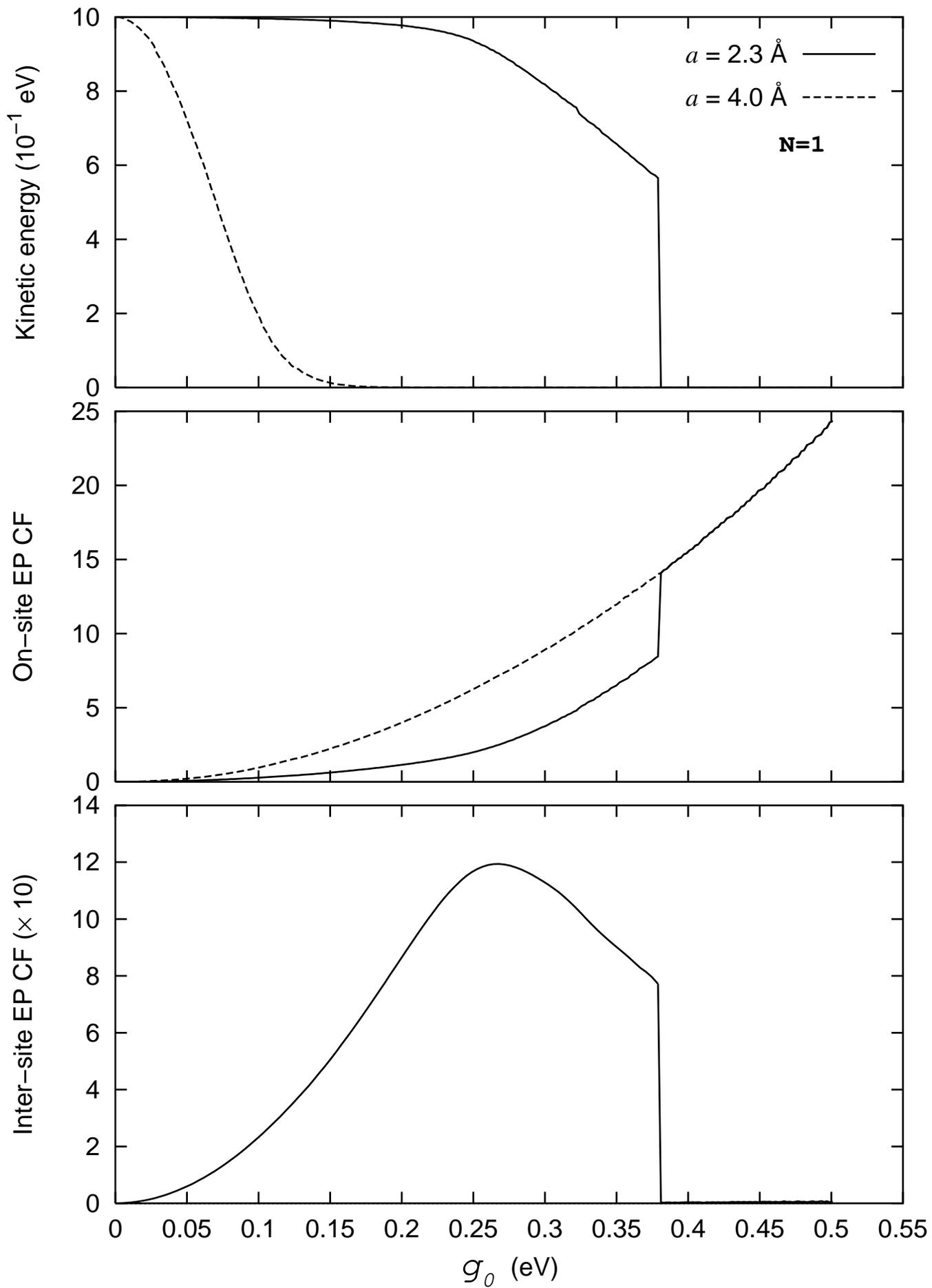
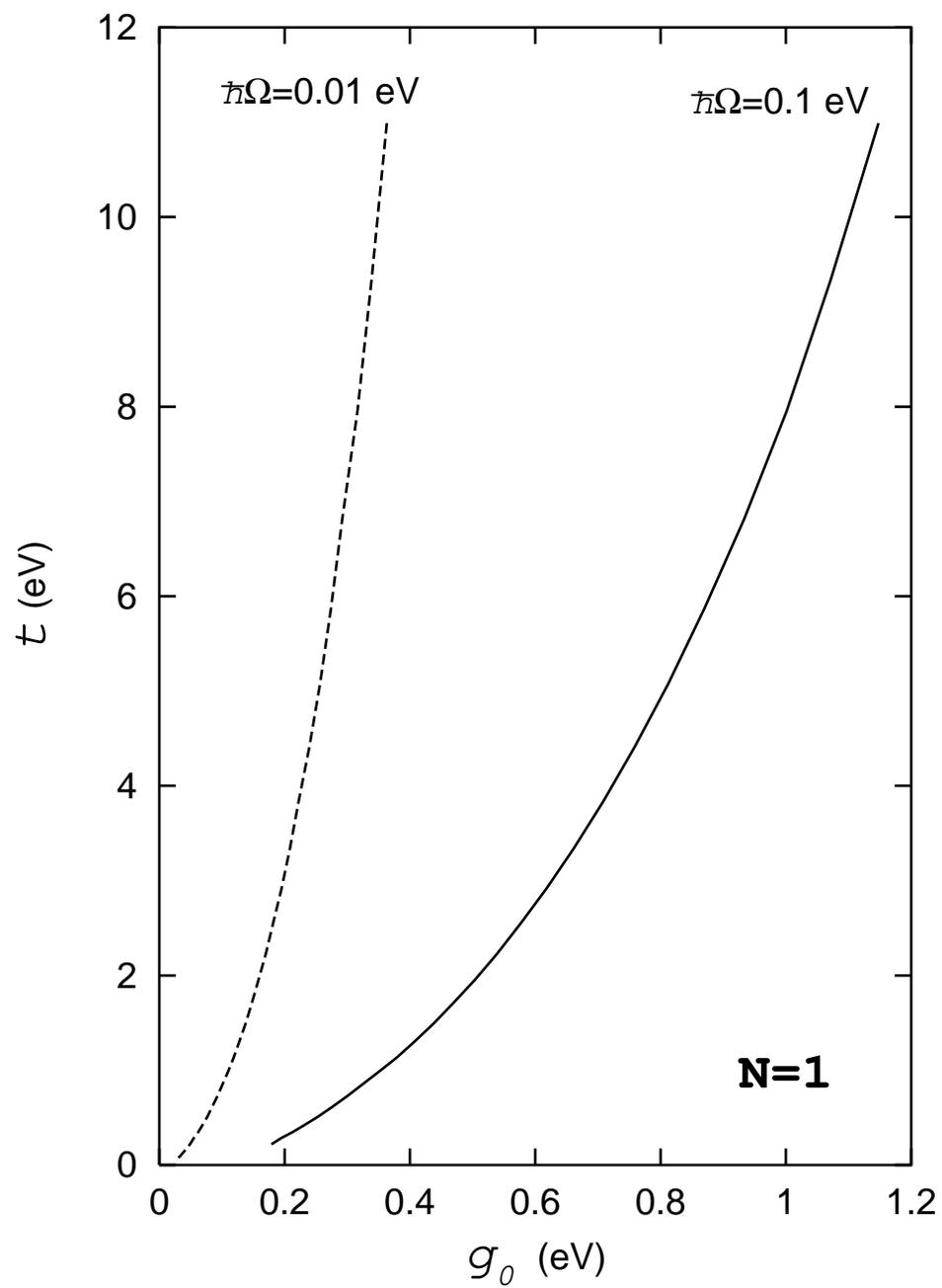
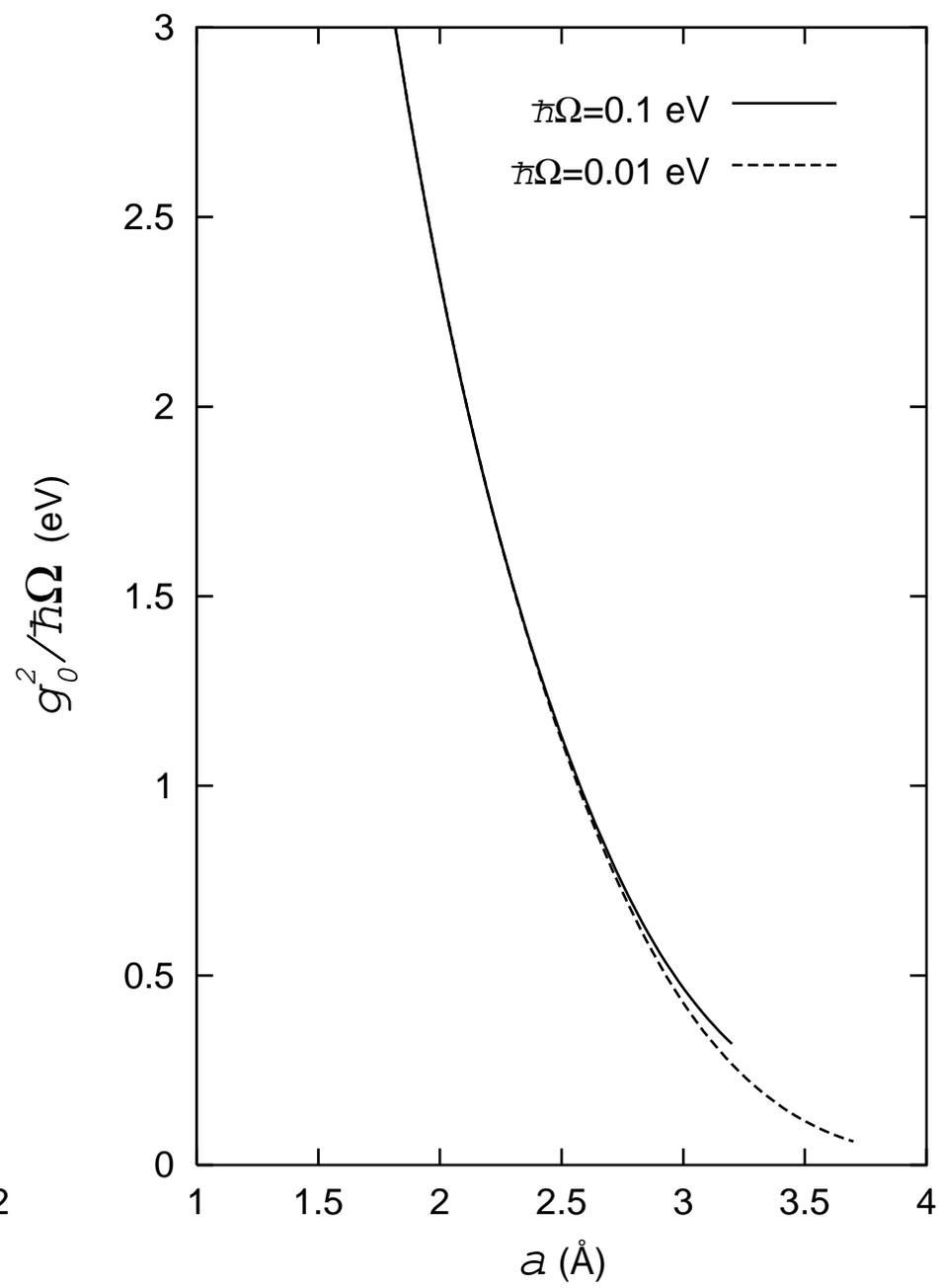
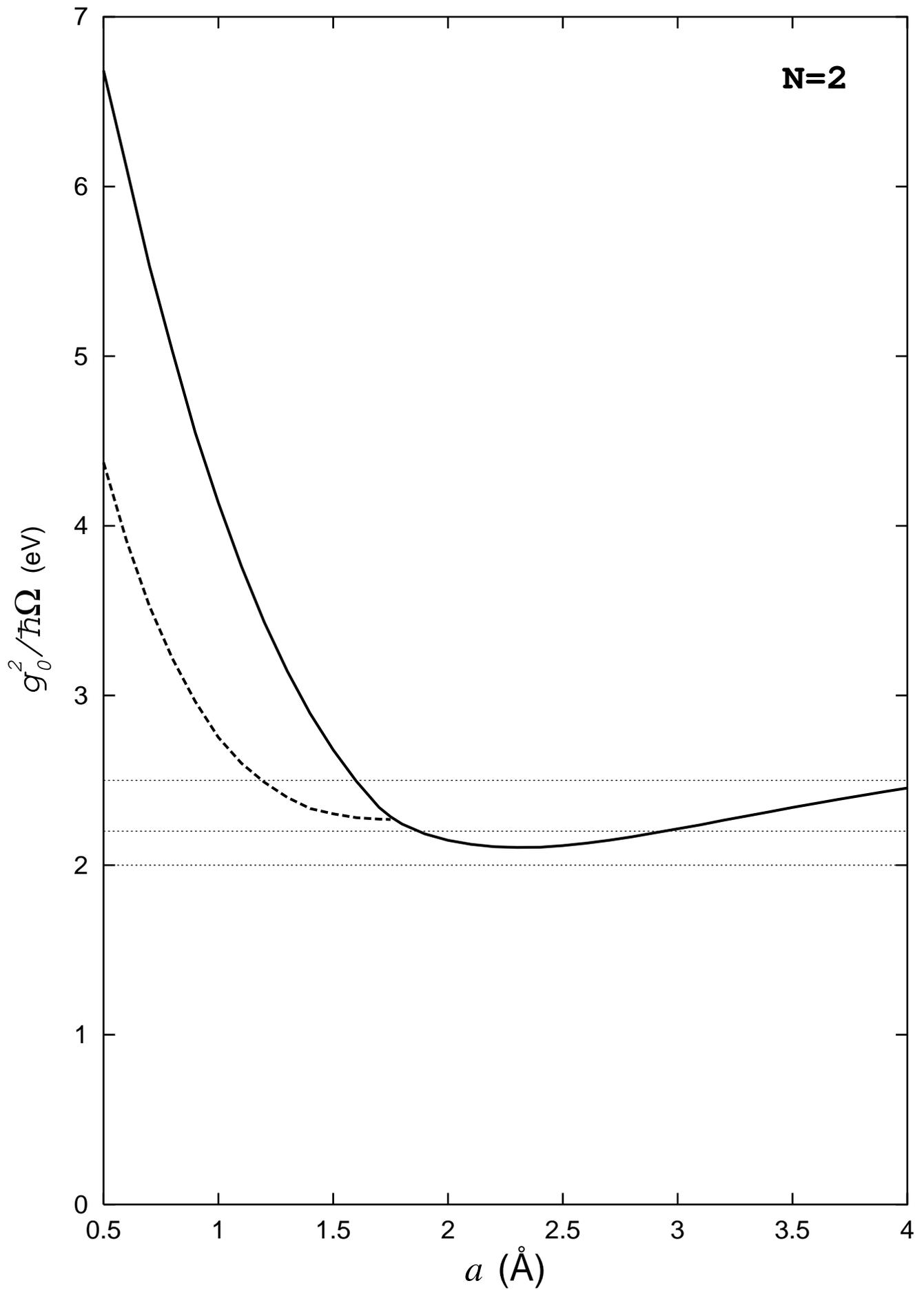
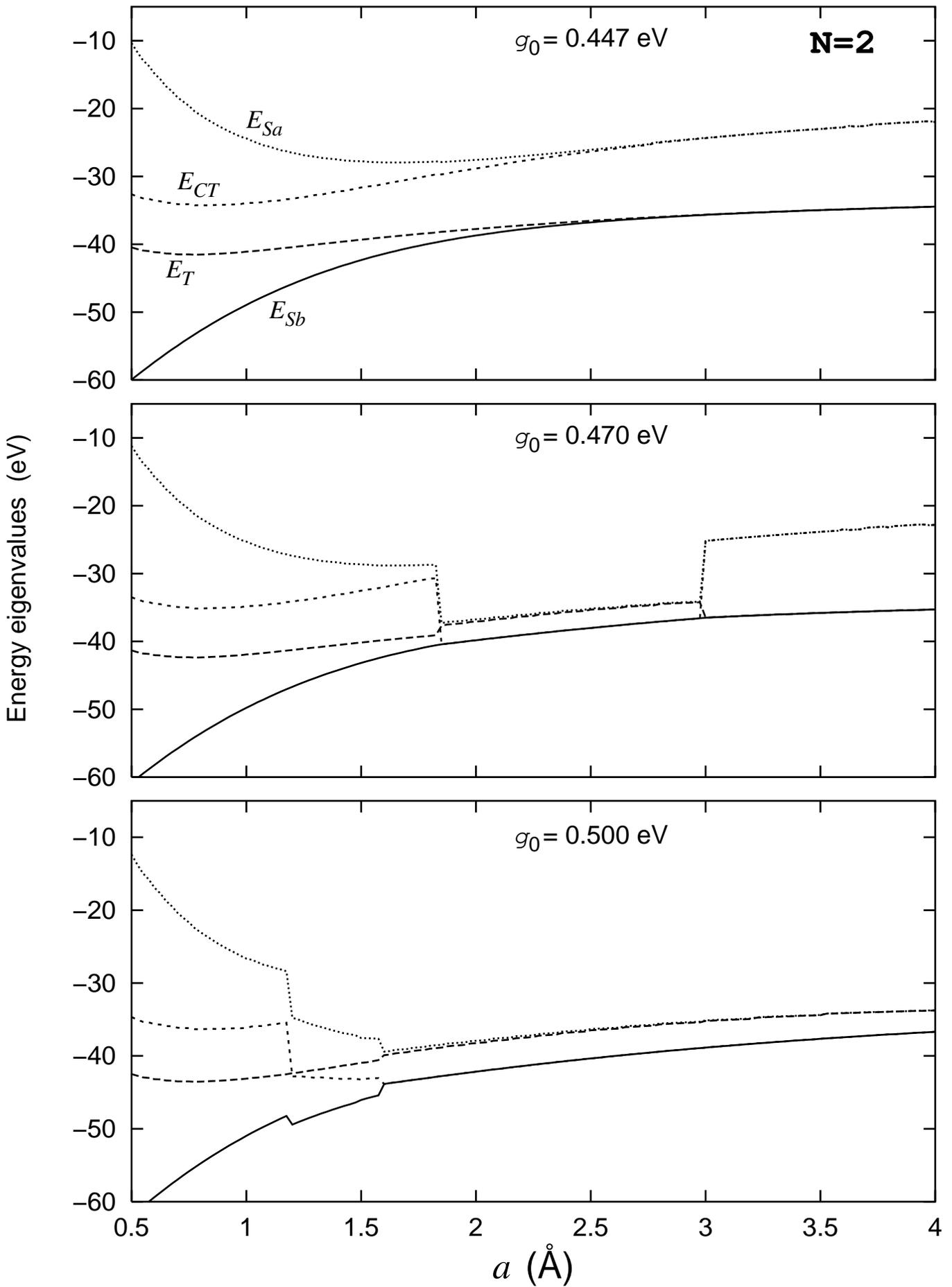
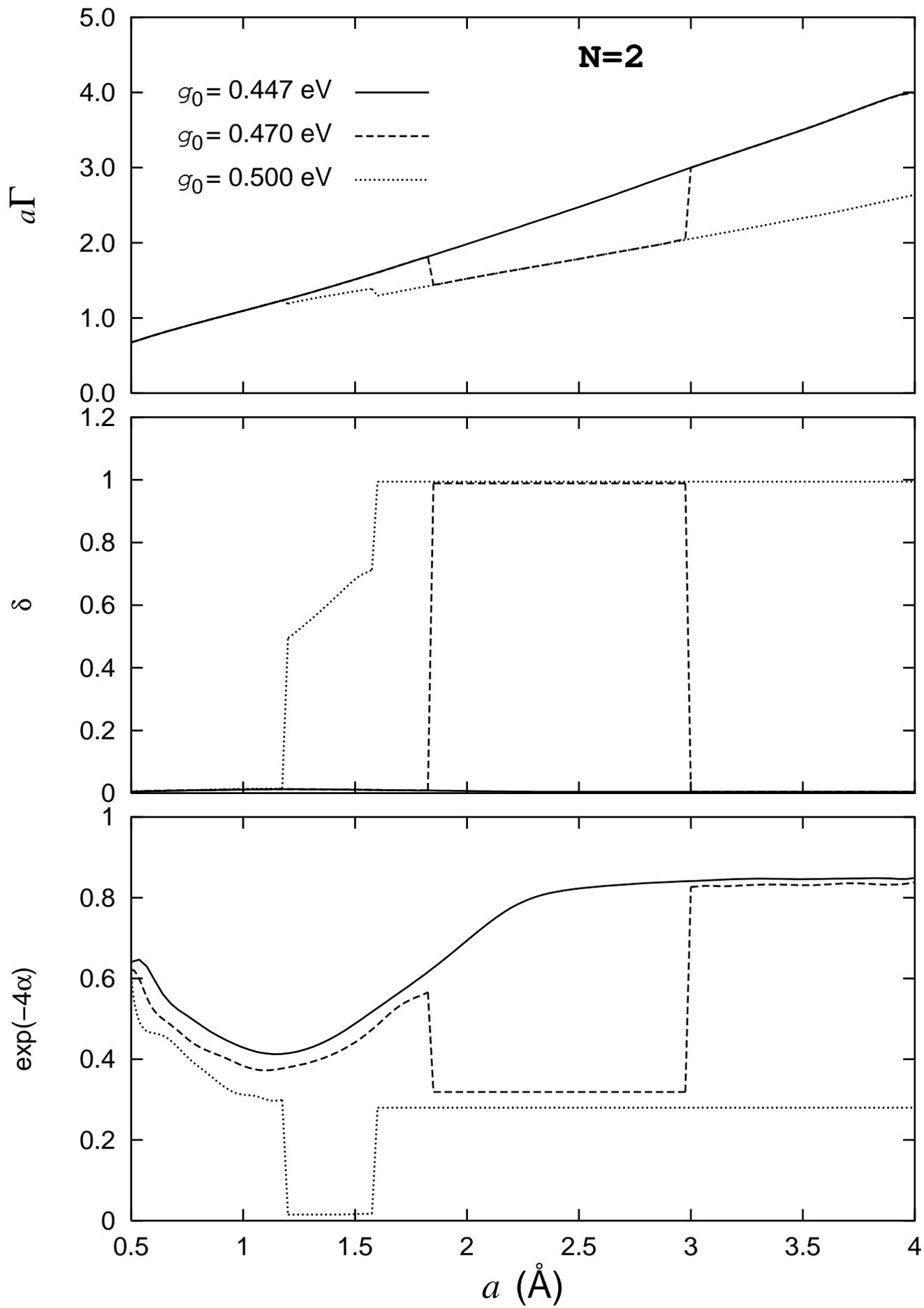
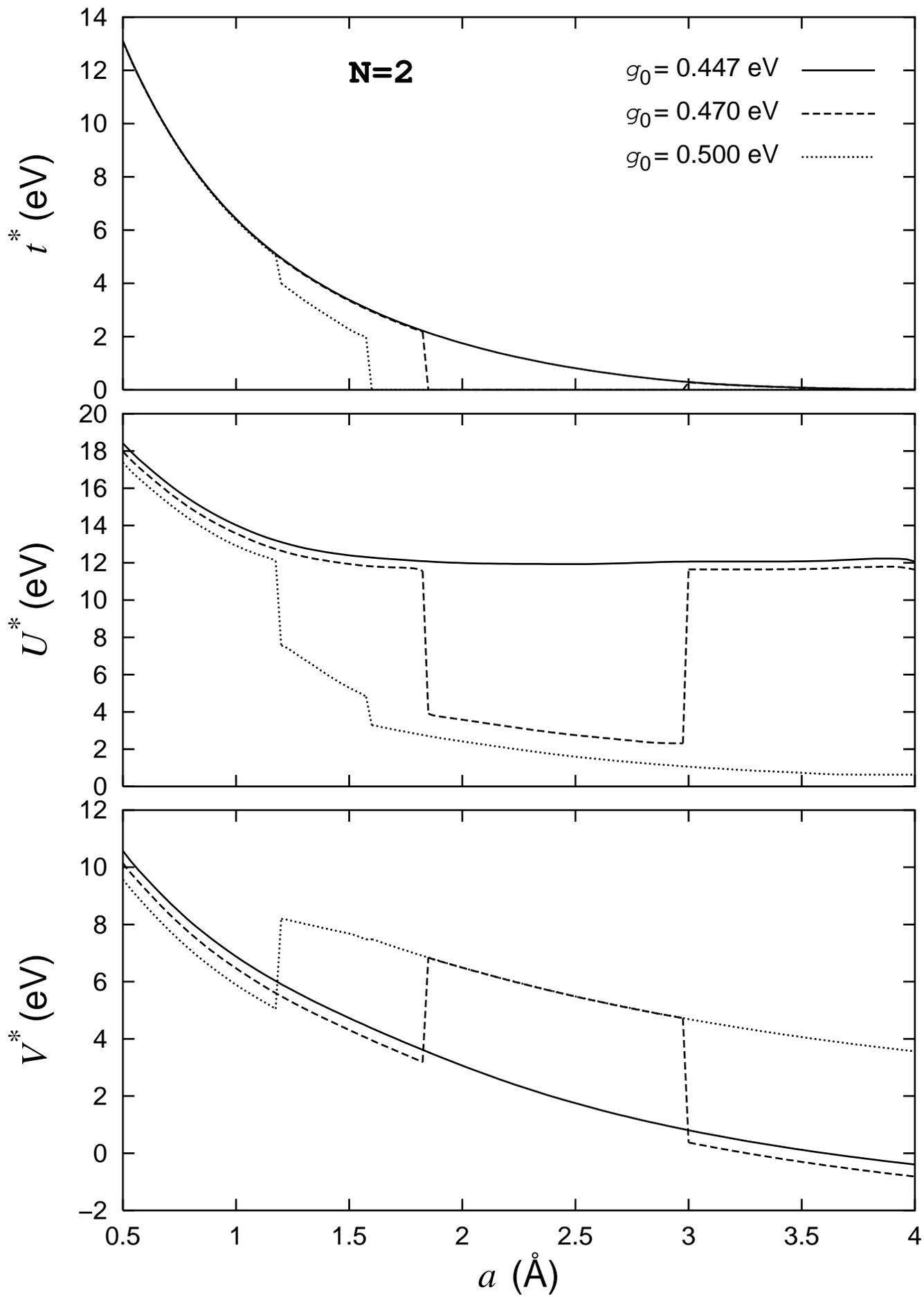
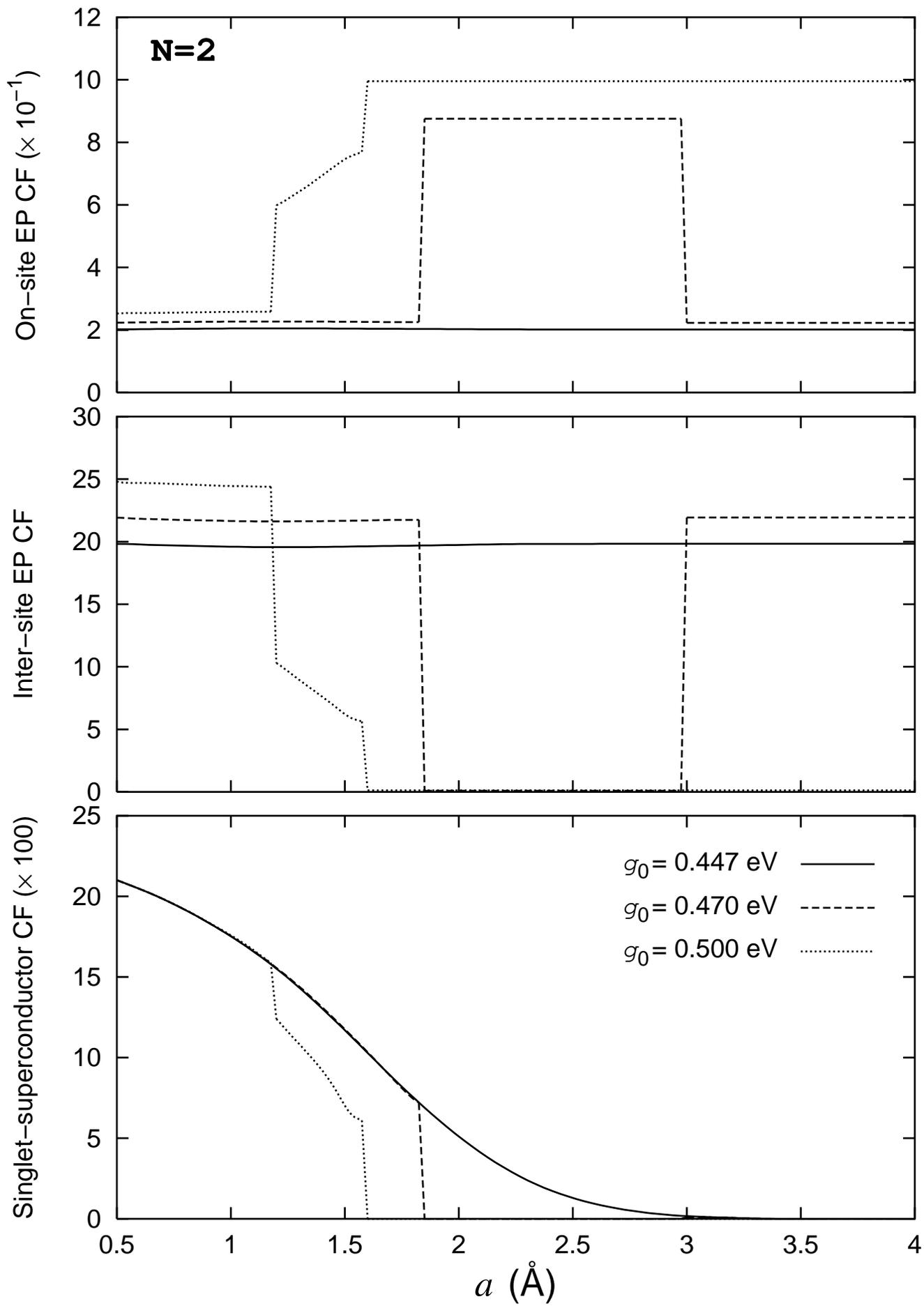
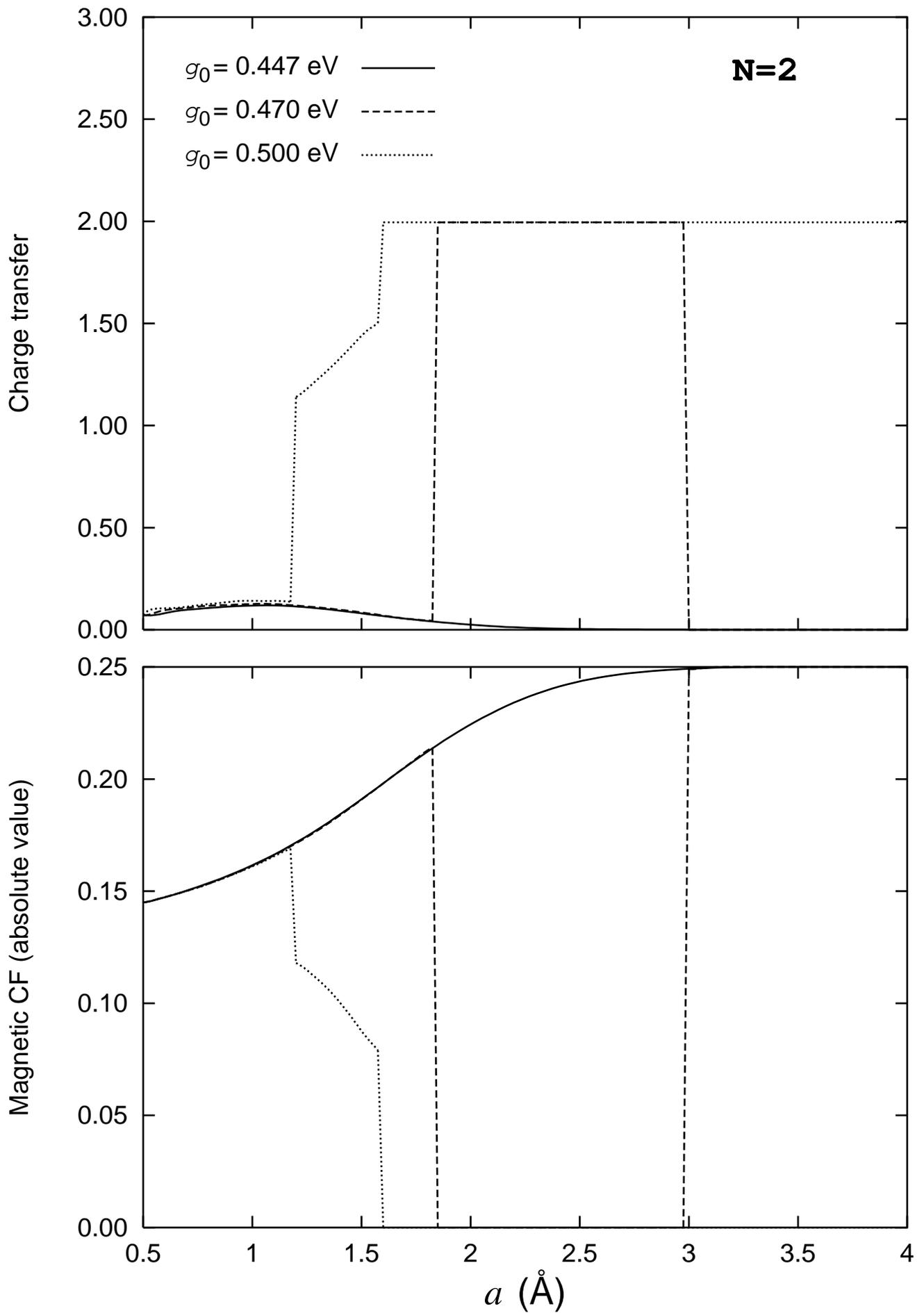
\end{document}